\documentclass[twocolumn,showpacs,preprintnumbers,amsmath,amssymb,prb]{revtex4}

\usepackage{graphicx}
\usepackage{dcolumn}
\usepackage{textcomp}
\usepackage{bm}

\def\rm#1{\mathrm{#1}}

\begin{document}


\title{Quantum-Dot Cellular Automata using Buried Dopants}

\author{Jared H. Cole}
 \email{j.cole@physics.unimelb.edu.au}
\author{Andrew D. Greentree}
\author{Cameron J. Wellard}
\author{Lloyd C. L. Hollenberg}
\author{Steven Prawer}
\affiliation{
Centre for Quantum Computer Technology, School of Physics, The University of Melbourne, Melbourne, Victoria 3010, Australia.
}%

\date{\today}

\begin{abstract}
The use of buried dopants to construct quantum-dot cellular automata is investigated as an alternative to conventional electronic devices for information transport and elementary computation.  This provides a limit in terms of miniaturisation for this type of system as each potential well is formed by a single dopant atom.  As an example, phosphorous donors in silicon are found to have good energy level separation with incoherent switching times of the order of microseconds.   However, we also illustrate the possibility of ultra-fast quantum coherent switching via adiabatic evolution.  The switching speeds are numerically calculated and found to be 10's of picoseconds or less for a single cell.  The effect of decoherence is also simulated in the form of a dephasing process and limits are estimated for operation with finite dephasing.  The advantages and limitations of this scheme over the more conventional quantum-dot based scheme are discussed.  The use of a buried donor cellular automata system is also discussed as an architecture for testing several aspects of buried donor based quantum computing schemes.
\end{abstract}

\maketitle

\section{Introduction}
The use of quantum systems to perform computing tasks is an area of continued interest, both in the context of fully coherent quantum computers and the more general area of nano-computing.  Recently there has been much interest in producing an experimental analogue of cellular automata at the micro or even nanometer scale\cite{Tougaw:94}\cite{Smith:99}.  This could provide an alternative architecture with which to build standard logic gates and information channels, which have low dissipation\cite{Timler:03} and fewer control gates compared to conventional transistor based logic.  Significant progress has been made in this area with the experimental demonstration of Quantum-Dot Cellular Automata (QDCA) cells constructed from Aluminium quantum-dots\cite{Orlov:97,Orlov:99} and GaAs/AlGaAs heterostructures\cite{Gardelis:03}.  Recently, the operation of the functional components of a QDCA logic gate has been demonstrated experimentally\cite{Amlani:99}.  This type of system has also been investigated as a possible architecture for quantum computing using QDCA qubits\cite{Toth:01} and more recently as a candidate for a decoherence free subspace\cite{Oi:04}.

In this paper we explore an alternative QDCA architecture using buried dopants in semiconductors, as this provides a very strongly confined potential and well characterised energy levels.  A recent proposal describes a charge-based qubit for quantum computing using phosphorous donors in silicon\cite{Hollenberg:04,Dzurak:03}.  We show that this treatment can be applied to a system of dopants arranged in the layout of a QDCA.  We refer to these structures as Buried Dopant Cellular Automata or BDCA.  Such a device could be fabricated by either direct atomic placement\cite{Schofield:03,Tucker:01} or ion-implantation\cite{Dzurak:03,Schenkel:03}.  Numerical estimates are given for the case of phosphorous donors in silicon but the concepts are generally applicable to other dopants.  This is of particular interest given the recent advances in single dopant placement and cluster based charge transfer experiments using phosphorous in silicon\cite{Clark:03}.

Conventional QDCA rely on \emph{incoherent} evolution (governed by the $T_1$ relaxation time) to mediate transitions between the logical states.  For phosphorous donors in silicon we estimate this relaxation time and find it to be of the order of microseconds.  We also investigate an alternative ultra-fast (picosecond) switching mechanism, namely \emph{coherent} evolution between defined system eigenstates.  This approach is central to the use of buried donors and is also applicable to coherent tunnelling between quantum dots or superconducting systems.  This constitutes an alternative evolution mechanism for QDCA schemes where coherence can be maintained long enough for the cell to be switched from one classical state to another without the need for long coherence times which are typically required for quantum computing applications.  The relevant time-scales and appropriate pulsing sequences with and without dephasing are discussed and the scaling behaviour of the system is investigated.

\section{Quantum-Dot Cellular Automata}
\begin{figure} [tb!]
\centering{\includegraphics[width=4cm]{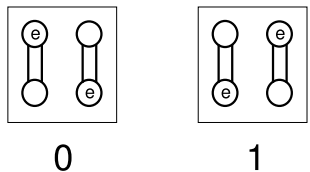}} \caption
{Two possible states for a basic QDCA cell where the 0 and 1 states constitute the ground or `computational' states and `e' labels the position of the electrons.\label{fig:QCA}}
\end{figure}

The simplest QDCA is a cell composed of four quantum-dots containing two mobile electrons which can move between the dots via tunnel junctions.  The electrons tend to occupy diagonally opposite sites to minimise the energy due to the Coulombic interaction.  These two ground (or computational) states are labelled zero and one, see Fig.~\ref{fig:QCA}, where `e' indicates the position of the electrons.  The next highest energetic states are non-computational states and ideally are only transiently populated during correct operation.  

\begin{figure} [tb!]
\centering{\includegraphics[width=6cm]{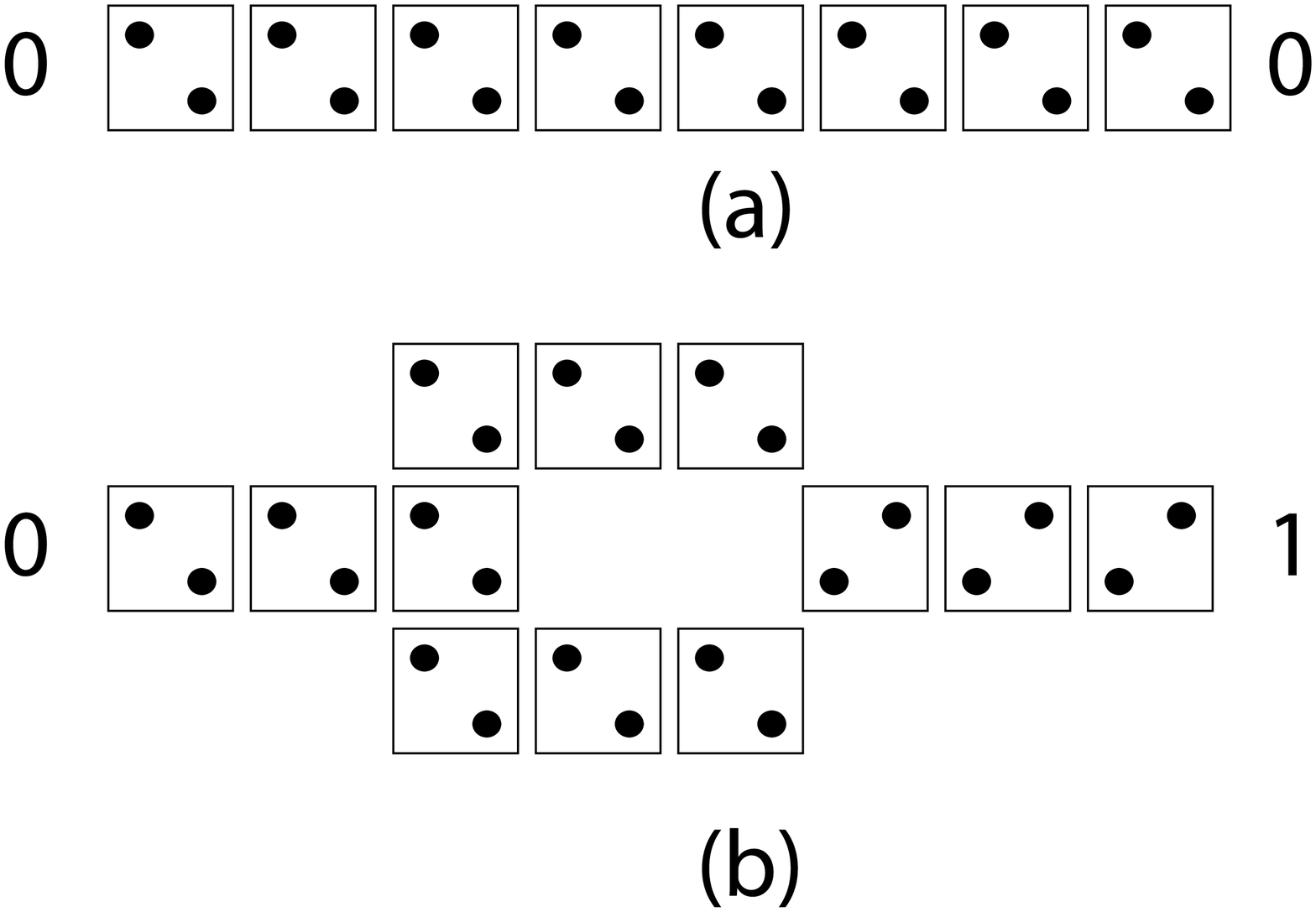}} \caption
{Layout for a QDCA wire (a) and inverter (b) which demonstrate information transfer and binary inversion respectively after Tougaw and Lent\cite{Tougaw:94}.\label{fig:QCAwire}}
\end{figure}

If two dots are placed next to each other, one cell influences the state of the other cell via capacitative coupling.  In an array of cells, when the first cell is switched from one computational state to the other, the rest of the chain relaxes to minimise the energy of the total system.  The result of this relaxation is to transfer the state information of the initial cell along the chain without net electron flow and minimal energy dissipation.  The speed at which this switching occurs is governed by the incoherent tunnelling rate of the junctions, the inverse of which is referred to as the $T_1$ or relaxation time.  Classical information processing can be performed in this scheme, as shown in Fig.~\ref{fig:QCAwire} for a QDCA wire and inverter\cite{Tougaw:94}.  It is also possible to realise non-trivial classical computation, such as a full-adder, using this scheme\cite{Snider:99b}.

\section{Buried Donors and the Hydrogenic Approximation}\label{sec:HydroApprox}
The use of buried donors in a semiconductor matrix has been discussed for charge-based quantum computing\cite{Hollenberg:04,Clark:03,Schenkel:03}.  While the advantages of semiconductor fabrication and gate control are well known, the fast dephasing and relaxation effects mean that charge-based quantum computing using buried donors is still technically difficult.  On the other hand the QDCA architecture is not as seriously affected by dephasing or relaxation, as the system is always in the ground state when it is measured.  We will investigate the possibility of using this concept for building QDCA based systems using phosphorous donors in silicon (Si:P), which we term a BDCA.  The basic idea is to construct an array of four ionised donors which contains two `free' electrons, therefore mimicking the layout of a conventional QDCA.  This also represents a limit in terms of miniaturisation for this form of nano-computing as each potential well is created by only one donor atom. 

In order to provide a model for the Si:P donor system, we will use the effective mass or `hydrogenic' approximation in which the outer shell electron of a phosphorous donor in silicon can be treated as a hydrogen orbital with the energies and distances scaled appropriately. A more complete treatment of effective mass theory for shallow donors is given by Kohn\cite{Kohn:55}.  Eq.~(\ref{eq:aeff})~and~(\ref{eq:Eeff}) give the scaling factors for the effective Bohr radius ($a^*_B$) and effective energy ($E^*$) of the donor electron in terms of the effective mass of the donor electron and the dielectric constant of the substrate,
\begin{equation}\label{eq:aeff}
a^{*}_{B}=\epsilon \frac{m_{e}}{m^{*}} a_{B},
\end{equation}
\begin{equation}\label{eq:Eeff}
E^{*}=\frac{m^{*}}{m_{e}}\frac{1}{\epsilon^2}E.
\end{equation}
The advantage of this approach is that solving a hydrogenic system with a small number of electrons is more tractable than a full electron calculation of the phosphorous donor within a silicon lattice.  These results also generalise to other shallow donor systems.  We will concentrate on phosphorous donors in silicon and therefore quote the appropriate energy levels, where the conversion is $13.6\mathrm{eV}$ in a hydrogenic system is approximately equal to $20\mathrm{meV}$ for Si:P.  The effective Bohr radius of the phosphorous electron is approximately $3\mathrm{nm}$ for an effective electron mass $m^*=0.2m_e$ and a dielectric constant $\epsilon=11.7$.

\section{Effective Hamiltonian}\label{sec:Heff}
\begin{figure} [tb!]
\centering{\includegraphics[width=4cm]{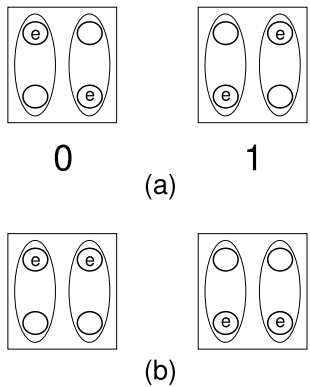}} \caption
{(a) The ground states of the buried donor BDCA cell where the positions of the electrons `e' are designated by top (T) and bottom (B).  These computational states are referred to as $|TB\rangle$ and $|BT\rangle$, and are assigned the logical values of 0 and 1 respectively.  (b) The excited or `non-computational' states are labelled $|TT\rangle$ and $|BB\rangle$ respectively and correspond to the first excited state of the system.\label{fig:BDQCAStates}}
\end{figure}

To provide a convenient formalism, we construct an effective Hamiltonian using the pseudo-spin approach to describe the BDCA system.  By defining each pair of phosphorous donors and their shared electron as a single pseudo-spin object we can define two states, top (T) and bottom (B), which specify the position of the electron.  Each BDCA cell then consists of a pair of these objects where the computational states are $|TB\rangle=|0\rangle$ and $|BT\rangle=|1\rangle$ respectively, as shown in Fig.~\ref{fig:BDQCAStates}(a).

In this way, the labelling for the non-computational states is $|TT\rangle$ and $|BB\rangle$, as shown in Fig.~\ref{fig:BDQCAStates}(b).  Initially we will assume that the electrons cannot move laterally so each electron is bound to its particular donor pair, as indicated by the ellipses in Fig.~\ref{fig:BDQCAStates}.  This would correspond to a situation where the direction of tunnelling is controlled by confining potentials or the geometry of the cell.  A more complete justification for this assumption is given later.

Once the position of the electrons is encoded using this pseudo-spin approach, an effective Hamiltonian can be developed using the Pauli spin representation,  
\begin{align}\label{eq:Heff}
H_{\mathrm{eff}} =& \epsilon_1\sigma^z_1+\epsilon_2\sigma^z_2+\Delta_1\sigma^x_1+\Delta_2\sigma^x_2 \nonumber \\
&+J_{\rm{xx}}\sigma^x_1\sigma^x_2+J_{\rm{yy}}\sigma^y_1\sigma^y_2+J_{\rm{zz}}\sigma^z_1\sigma^z_2 \nonumber \\
&+J_{\rm{xz}}\sigma^x_1\sigma^z_2+J_{\rm{zx}}\sigma^z_1\sigma^x_2, 
\end{align}
where the usual Pauli matrices are defined as 
\begin{eqnarray}
\sigma^x=\left[\begin{array}{cc} 
                    0 & 1 \\ 
                    1 & 0 
                    \end{array}\right],  
&  \sigma^y=\left[\begin{array}{cc} 
                    0 & -i \\ 
                    i & 0 
                    \end{array}\right],
&  \sigma^z=\left[\begin{array}{cc} 
                    1 & 0 \\ 
                    0 & -1 
                    \end{array}\right].  \nonumber  
\end{eqnarray}\label{eq:Paulis}
The notation $\sigma_i$ is used to refer to the matrix applied to the $i\mathrm{th}$ donor pair and the coefficients $\epsilon$, $\Delta$ and $J$ are determined as follows.

The true hydrogenic Hamiltonian\cite{Slater:63} of the two electron/four donor system is given (in scaled atomic units) by 
\begin{align}\label{eq:TrueH}
H=&-\triangledown_1^2-\triangledown_2^2-2(\frac{1}{r_{\rm{1a}}}+\frac{1}{r_{\rm{2a}}}+\frac{1}{r_{\rm{1b}}}+\frac{1}{r_{\rm{2b}}} \nonumber \\ 
&+\frac{1}{r_{\rm{1c}}}+\frac{1}{r_{\rm{2c}}}+\frac{1}{r_{\rm{1d}}}+\frac{1}{r_{\rm{2d}}}-\frac{1}{r_{\rm{12}}}),
\end{align}
where $r_{ij}$ is the separation between the $i\rm{th}$ electron and the $j\rm{th}$ donor ($j=a,b,c,d$) and $r_{12}$ is the separation between the electrons.  Numerically evaluating this Hamiltonian within the basis of four states ($|TB\rangle$, $|BT\rangle$, $|TT\rangle$ and $|BB\rangle$) enables the elements of the matrix 
\begin{equation}\label{eq:Hij}
H_{ij}=\langle \psi_i|H|\psi_j \rangle
\end{equation}
to be found.  The elements of Eq.~(\ref{eq:Hij}) are then equated to the coefficients in Eq.~(\ref{eq:Heff}) to determine an effective Hamiltonian.  
The basis of states is represented using a linear combination of atomic orbitals (LCAO).  We use the anti-symmetric spatial wavefunction for the $H_2$ molecule as our basis wavefunction.  The spin-orbit coupling for donor electrons in silicon is known to be very small\cite{Appel:64} and therefore spin effects may be neglected as the spin and charge degrees of freedom are assumed to be separable at all times.  The elements of Eq.~(\ref{eq:Hij}) are then used to obtain estimates for the numerical coefficients in the effective hamiltonian ($H_{\rm{eff}}$) for a square BDCA cell of side length $R$.  

\begin{figure} [tb!]
\centering{\includegraphics[width=8.6cm]{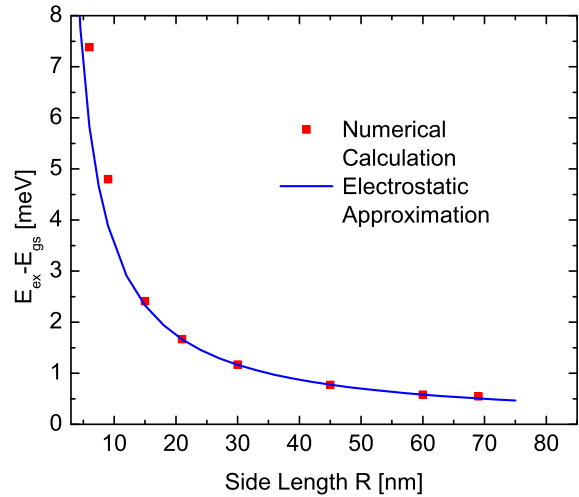}} \caption
{Energy difference between the ground ($E_{\rm{gs}}$) and first excited ($E_{\rm{ex}}$) states of a square BDCA cell made up of phosphorous donors in silicon.  The energy difference is computed for various donor separations (R) by numerically integrating the schrodinger equation. The points are full quantum mechanical calculations using the LCAO approach and the solid line is the energy difference determined analytically from simple electrostatic arguments, Eq.~(\ref{eq:G2E}).\label{fig:Ediff}} 
\end{figure}

We estimate the energy difference ($E_{\rm{ex}}-E_{\rm{gs}}$) between the excited (non-computational) and ground (computational) states of one cell as a function of system size.  This energy gap gives an estimate of the temperature at which the system must be operated to ensure that the non-computational states are not thermally excited.  The energy difference for a range of separations is plotted in Fig.~\ref{fig:Ediff}.  The points are found using the LCAO approach while the line is an approximation found using electrostatic arguments based on the geometry of the system and has the form 
\begin{equation}\label{eq:G2E}
E_{\rm{ex}}-E_{\rm{gs}}=\frac{(2-\sqrt{2})E^{*}a^{*}_{B}}{R},
\end{equation}
where $R$ is the side length of the BDCA cell in $\mathrm{nm}$ and $a^{*}_{B}=3\rm{nm}$, $E^{*}=20\rm{meV}$ are the effective Bohr radius and effective ground state energy respectively.  This approximation is found to be valid in the region where the electron wavefunctions do not strongly overlap with each other but deviates from the numerical results for $R\lesssim 10\mathrm{nm}$. 

Assuming a Boltzmann distribution for the energy states at finite temperature, the occupation probability ($P_{ex}$) of the excited states ($|TT\rangle$ and $|BB\rangle$) is given by 
\begin{equation}\label{eq:GndOccvsT}
P_{ex}\approx e^{-\frac{\Delta E}{k_BT}},
\end{equation}
where $E_{\rm{ex}}-E_{\rm{gs}}=\Delta E > 0$ is the energy difference between the ground and excited states.  For a separation of $\mathrm{15nm}$, we require an operating temperature of $<\!3\rm{K}$ to achieve $>\!99\%$ occupation of the computational states.  For an operating temperature of $100\rm{mK}$, the occupancy of the computational states is approximately $100\%$.

For the rest of the discussion, we will use the numerical coefficients calculated for a square cell of side length $15\mathrm{nm}$.  As this is equivalent to a cell size of approximately $5$ Bohr radii, the electrons can be said to be well localised and overlap effects are not significant.  In order to simplify the situation, we will also set the $\sigma^x$, $\sigma^x\sigma^x$ and $\sigma^y\sigma^y$ type terms to zero and then reintroduce them later in a systematic fashion.  This corresponds to a situation where a surface gate potential is used to control overlap of the electron wavefunctions and therefore control the tunnelling rate between donors.  For large cells sizes, this is a good approximation to the physical situation as the tunnelling rate without an applied barrier bias will be very low.  For small cell sizes there will be wavefunction overlap even without a barrier gate.  In this case a confining potential can be used to localise the electrons and provide more control over the tunnelling characteristics.  Given these approximations, many of the terms in Eq.~(\ref{eq:Heff}) are approximately zero and the only significant term for $R=15\rm{nm}$ is $J_{\rm{zz}}=1.21$ $\mathrm{meV}$ which is due to the Coulombic repulsion between the electrons.

\section{BDCA Switching}
\begin{figure} [tb!]
\centering{\includegraphics[width=8.6cm]{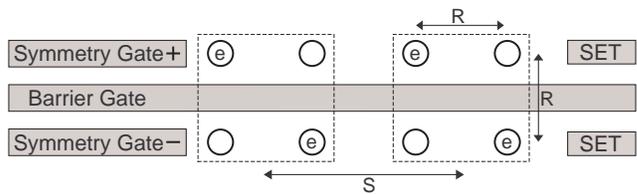}} \caption
{Simplified layout of a BDCA chain, where the cell size is labelled $R$ and the cell spacing ($S$) is the distance between the centre of neighbouring cells.  The symmetry and barrier surface gates are used to perform switching and control the tunnelling rate respectively.  The circles represent the position of the donors.  The position of the electrons `e' is shown for the ground state configuration which corresponds to some non-zero bias on the control gates with the labelled polarity.  The position of the electrons are measured with single-electron transistors (SET).\label{fig:QCCA_layout}}
\end{figure}

To study how a BDCA cell would switch from one computational state to another, we consider the effect of control gates and nearby cells.  We propose a structure where pairs of donors are positioned in a line with a surface `barrier' gate constructed between them which is used to control the tunnelling rate between the pairs of donors.  At the end of the chain, `symmetry' gates allow the system to be switched from one state to another, depending on the bias applied.  Sensitive electrometers, such as single-electron transistors (SET)\cite{Grabert:92}, are used to measure the position of the electron at the other end of the chain.  A diagram of this concept is shown in Fig.~\ref{fig:QCCA_layout}.

Most of the following section refers to a single BDCA cell made from two pairs of donors for simplicity, but the discussion applies equally to a long chain of donor pairs.  The effect of the barrier (B) and symmetry gates (S) for this cell can be included by adding terms to the effective Hamiltonian, Eq.~(\ref{eq:Heff}), of the form
\begin{equation}\label{eq:Controls}
E_S(t) \sigma^z_1 - E_B(t) (\sigma^x_1 + \sigma^x_2),
\end{equation}
where the magnitude of the coefficients ($E_S$ and $E_B$) are controlled by the voltages applied to the gates.  

The symmetry gates localise the system in one of the two computational states, based on the gate's polarity.  The voltages applied to each symmetry gate have equal magnitude but opposite sign to ensure a symmetrical effect on the chain.  

The barrier gate controls the tunnelling rate between pairs of donors by repelling or attracting the electron clouds.  If there is a large enough separation between donors, the barrier gate is required to allow tunnelling by reducing the potential barrier that the electron feels.  This justifies our initial assumption that when the separation is large enough tunnelling only occurs between donor pairs and not along the BDCA chain.  The use of compensating gates could also be used to confine the electrons and therefore prevent tunnelling in unwanted directions.  

The effect of these surface gates has been modelled as pure $\sigma^x$ and $\sigma^z$ terms in the Hamiltonian due to the symmetry of the system.  We estimate the surface gate voltages to be $100-1000\rm{mV}$ depending on the donor depth and the presence of an oxide barrier layer, based on estimated gate voltages for charge-qubits\cite{Hollenberg:04,Lee:04}.
Assuming arbitrary high precision in donor placement, the barrier gate has an equal effect on donors either side of the barrier and can be considered a pure $\sigma^x$ gate.  The barrier gate would also induce $\sigma^x\sigma^x$ and $\sigma^y\sigma^y$ style coupling but this is expected to be small compared to the $J_{\rm{ZZ}}$ and pure $\sigma^x$ coupling. In addition, the barrier gate can also have a negative bias applied to improve the localisation of the computational states during readout, though this is not directly modelled here.  
While the symmetry gates would not be pure $\sigma^z$ (having some residual $\sigma^x$ effect due to wave-function overlap) this could also be compensated for by using the barrier gate or additional compensation gates. 

\begin{figure} [tb!]
\centering{\includegraphics[width=8.6cm]{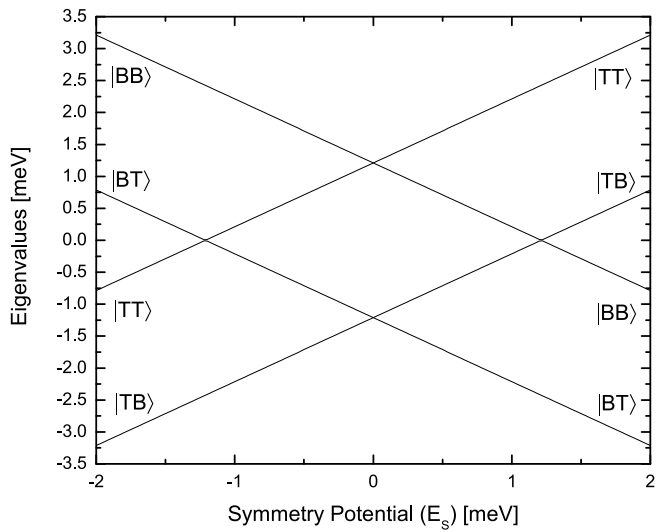}} \caption
{Eigenspectrum for a four donor cell as the symmetry potential ($E_S$) is swept from -2 to 2 (meV) with no applied barrier potential ($E_B=0$).\label{fig:ex_000}}
\end{figure}

In Fig.~\ref{fig:ex_000} the eigenspectrum for a single BDCA cell is plotted as a function of the symmetry gate potential ($E_S$) for zero barrier gate potential.  The computational states ($|TB\rangle$ and $|BT\rangle$) are localised even for a very small symmetry potential.  Also note that the two computational states are degenerate when there is no potential difference applied to the control gates, as expected.

\section{Incoherent Switching}
If we allow the system to evolve via incoherent relaxation (in direct analogy with the quantum-dot schemes), the transition from a high to a low energy eigenstate of the system is mediated by phonons in the lattice.  For the moment we will ignore the effect of the barrier gate and assume the electrons are well localised.  This situation is valid over time scales greater than the dephasing time of the donor wavefunction or where the barrier gate has a negative bias producing a high potential barrier.  To obtain an estimate for this rate, we assume that the electrons within a BDCA cell can relax independently and we therefore use a similar approach to that used to estimate the relaxation rate for charge-qubits based on buried donors\cite{Barrett:03,Fedichkin:04}.  Following the approach used by Barrett and Milburn\cite{Barrett:03} and Bockelmann and Bastard\cite{Bockelmann:90}, we write the relaxation rate due to thermal phonons,
\begin{equation}\label{eq:Gammaph}
\Gamma_{\mathrm{ph}}=\frac{64D^2q_{if}^3[n_B(E,T_{\mathrm{ph}})+\alpha][1-\mathrm{sinc}(q_{if}R)]}{\pi\rho\hbar c_s^2[(q_{if}a_B)^2+4]^4},
\end{equation}
where $\alpha=1$ for emission and $0$ for absorption of a phonon, $q_{if}$ is the wavenumber of a phonon with a magnitude equal to the energy difference between the states ($|q_{if}|=E/\hbar c_s$), $R$ is the separation between the donors and $n_B(E,T_{\mathrm{ph}})=[\exp{(E/k_BT_{\mathrm{ph}})}-1]^{-1}$ is the Bose occupation function for a bath of phonons at temperature $T_{\mathrm{ph}}$.  We have ignored effects due to coherent tunnelling and used the following parameters\cite{Barrett:03} for Si:P where $D=3.3\mathrm{eV}$ is the deformation potential, $\rho=2329\mathrm{kg m^{-3}}$ is the density of silicon, $c_s=9.0\times 10^3\mathrm{ms^{-1}}$ is the speed of sound in silicon and $a_B=3\mathrm{nm}$ is the effective Bohr radius of the donor electron.

\begin{figure} [tb!]
\centering{\includegraphics[width=6cm]{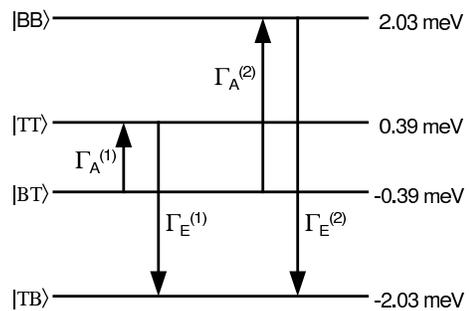}} \caption
{Energy level diagram for a single BDCA cell in the presence of a well localised neighbouring cell, where the cell size $R=15\mathrm{nm}$ and there is a cell spacing $S=30\mathrm{nm}$.  The direct transition ($|BT\rangle \rightarrow |TB\rangle$) is suppressed as the interaction is phonon mediated and must therefore proceed via single-electron transitions.  The two (first order) decay paths from one computational state to the other are illustrated with their associated transition rates for absorption ($A$) and emission ($E$).\label{fig:EnergyLevels}} 
\end{figure}

To estimate the incoherent switching time, we calculate the energy levels for a BDCA cell (the target cell) assuming that a neighbouring cell is well localised.  In the incoherent limit, we can assume that each electron is well localised and so the effect of the neighbour cell is to lift the degeneracy of the computational states of the target cell.  For a cell of side length $R=15\mathrm{nm}$ and cell spacing $S=30\mathrm{nm}$, a well localised neighbour induces an energy splitting of $1.64\mathrm{meV}$ between the computational states of the target cell.  This is calculated using the difference in electrostatic repulsion between the target cell and its neighbour.  The splitting caused by the neighbouring cell can be modelled as a bias on the symmetry gate of the target cell.  In this case the equivalent symmetry gate bias is $E_S=0.82\mathrm{meV}$.  The resulting energy levels can be read from Fig.~\ref{fig:ex_000}.  Using these energy levels we can calculate the relaxation rate from the first excited state to the ground state ($|BT\rangle \rightarrow |TB\rangle$) and estimate the switching time of the system.  Relaxation in this system is phonon mediated and therefore acts on each electron separately.  This means the direct transition from state $|BT\rangle$ to state $|TB\rangle$ is suppressed and instead the system must relax via a co-tunnelling (two electron) process which requires absorption and emission of phonons to reach the ground state.  The two possible (first order) decay paths are illustrated in Fig.~\ref{fig:EnergyLevels} for the energy levels corresponding to a symmetry potential of $0.82\mathrm{meV}$.

\begin{figure} [tb!]
\centering{\includegraphics[width=8.6cm]{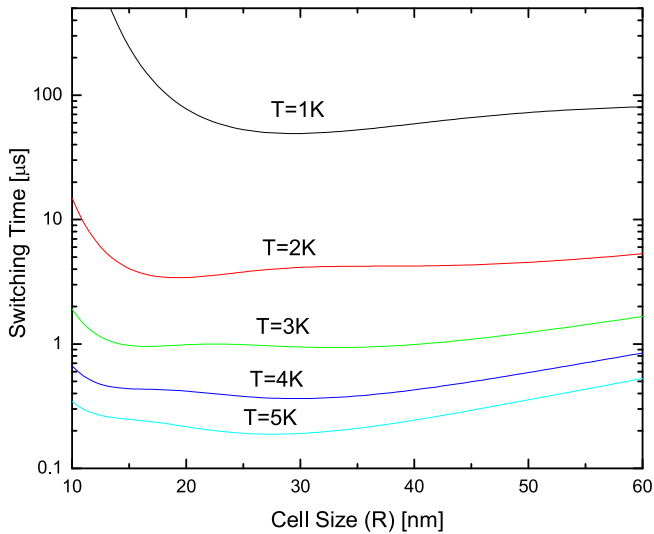}} \caption
{Incoherent switching time calculated for a range of cell sizes ($R$) with the spacing between the neighbouring cells given by the cell centre-to-centre distance $S=2R$.  Higher operating temperatures result in faster switching times but also result in higher excited state populations, reducing the overall readout fidelity.\label{fig:IncoherentSwitching}} 
\end{figure}

To estimate the total relaxation rate, we add the co-tunnelling rates\cite{Grabert:92} for the two possible paths, 
\begin{equation}\label{eq:cotunnelling}
\Gamma_{\rm{Total}}=\frac{\Gamma_A^{(1)}\Gamma_E^{(1)}\hbar}{|E_{|TT\rangle}-E_{|BT\rangle}|}+\frac{\Gamma_A^{(2)}\Gamma_E^{(2)}\hbar}{|E_{|BB\rangle}-E_{|BT\rangle}|},
\end{equation}
where $\Gamma^{(1,2)}_{A,E}$ are defined in Fig.~\ref{fig:EnergyLevels} and $E_{|k\rangle}$ is the energy of the $|k\rangle \mathrm{th}$ state.  For these energy levels and an operating temperature of $3\rm{K}$, the calculated relaxation rate $\Gamma_{\rm{Total}}=1.1\mathrm{MHz}$, which gives a switching time of $0.9 \mu s$.  This is almost two orders of magnitude slower than the estimated maximum switching rates of $90\mathrm{MHz}$ for Al/Al-oxide QDCA structures\cite{Orlov:99} at $70\mathrm{mK}$.  This is to be expected as there are no defined tunnel junctions in the buried donor case.  
The switching time is shown in Fig.~\ref{fig:IncoherentSwitching} for a range of operating temperatures and cell sizes ($R$) with the spacing between neighbouring cells given by the cell centre-to-centre distance $S=2R$.  While the switching rate does vary with cell size, the temperature effects dominate in this regime as the system requires enough thermal energy to mediate the two-electron transition.  At higher temperatures, faster switching is expected but the occupation of the ground state is reduced, according to Eq.~(\ref{eq:GndOccvsT}), resulting in an overall loss of fidelity.  

While the switching time is slow compared to modern microelectronics, it does demonstrate that if the system is initially setup in some excited state it will decay to the ground state.  This process could be used to intialise the system in a known state.

\section{Coherent Switching}\label{sec:CoherentEvol}
An alternative to the \emph{incoherent} switching is to use the high tunnelling rates of charge-based quantum-computing schemes to perform \emph{coherent} switching of the BDCA chain.  As we have relatively strong coupling between electrons in this system, we can consider adiabatic evolution as a mechanism to switch from one computational state to the other.  This is in exact analogy with the technique of Rapid Adiabatic Passage for electromagnetically induced population transfer of atoms and molecules\cite{Vitanov:01}.  While this method has similar advantages to the adiabatic switching discussed by Lent and Tougaw\cite{Lent:97}, the method discussed here is entirely coherent and therefore doesn't rely on dissipation to ensure the ground state is always occupied.

\begin{figure} [tb!]
\centering{\includegraphics[width=8.6cm]{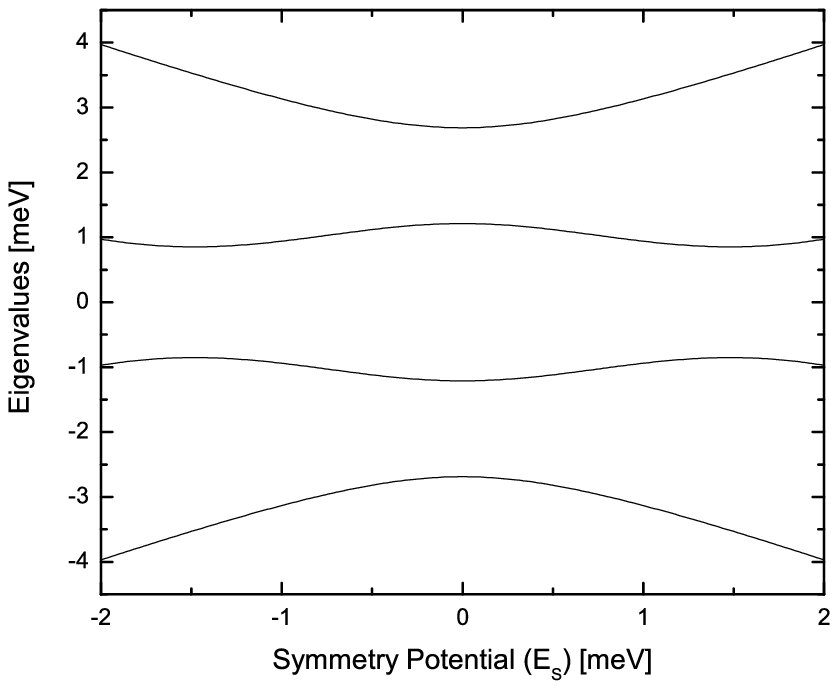}} \caption
{Eigenspectrum for a four donor cell as the symmetry potential ($E_S$) is swept from -2 to 2 (meV) with an applied barrier potential ($E_B$) of $1.2\mathrm{meV}$\label{fig:ex_003}}
\end{figure}

The effect of the barrier potential on the eigenspectrum of a BDCA cell is shown in Fig.~\ref{fig:ex_003}, this time with an applied barrier potential of $1.2 \mathrm{meV}$.  The effect of this is to lower the barrier within each donor pair, delocalising the electron and increasing the coupling between the donors.  For a barrier potential of $1.2 \mathrm{meV}$ there is now an energy gap between the ground and first excited states at the point of zero symmetry potential.  At this point the eigenstates include contributions from all four basis states, not just the computational states.  

When $E_S=S_{\rm{max}}$ and $E_B=1.2 \rm{meV}$ the computational ground state population has been reduced to $80\%$ (compared to without barrier gate induced coupling between the donors).  For example, when $E_S=-2\rm{meV}$ the lowest energy eigenstate is $0.39|TT\rangle+0.89|TB\rangle+0.14|BT\rangle+0.17|BB\rangle$ whereas for $E_S=2\rm{meV}$ it is $0.17|TT\rangle+0.14|TB\rangle+0.89|BT\rangle+0.39|BB\rangle$ when written out in the position basis.

\begin{figure} [tb!]
\centering{\includegraphics[width=8.6cm]{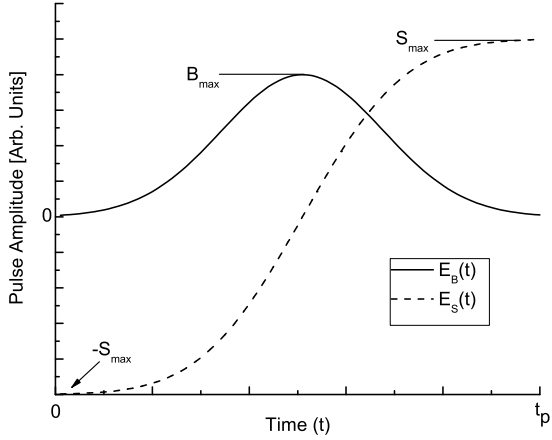}} \caption
{The symmetry ($E_S$) and splitting ($E_B$) potentials which are applied to the system to achieve adiabatic evolution, where $t_p$ is the time over which the pulse is applied and the standard deviation of the pulses ($\sigma$) is set to $t_p/6$.  $B_{\mathrm{max}}$ and $S_{\mathrm{max}}$ are the maximum barrier and symmetry potentials and are used to control the amount of tunnelling and localisation respectively.\label{fig:pulse_seq}}
\end{figure}

The energy gap between the ground and first excited state allows adiabatic evolution to be used to shift the population from one computational state to the other, provided that the adiabatic criteria is satisfied.  The adiabatic criteria can be stated as\cite{Messiah:65}
\begin{equation}\label{eq:AdiabaticEq}
\frac{\langle e|\frac{\partial H}{\partial t}|g\rangle}{\left|\langle e|H|e\rangle-\langle g|H|g\rangle\right|^2} \ll 1,
\end{equation}
where $|g\rangle$ and $|e\rangle$ are the ground and first excited states respectively.
The pulse scheme given in Fig.~\ref{fig:pulse_seq} can be applied in order to achieve the energy level splitting while still ensuring the computational states are highly populated for readout.  This involves applying a Gaussian pulse to the barrier gate while simultaneously switching the control gates from one polarity to the other.  The following functions were used for the barrier and symmetry gate potentials
\begin{equation}\label{eq:EB}
E_B(t)=B_{\mathrm{max}}\exp\left[-\frac{(t-t_p/2)^2}{2\sigma^2}\right],
\end{equation}
\begin{equation}\label{eq:ES}
E_S(t)=S_{\mathrm{max}}\mathrm{erf}\left[-\frac{(t-t_p/2)}{\sqrt{2}\sigma}\right],
\end{equation} 
where $t_p$ is the total time over which the pulse sequence is applied and $\sigma$ is the standard deviation of the pulse, which was set to $\sigma=t_p/6$.  $B_{\rm{max}}$ and $S_{\rm{max}}$ are the maximum barrier and symmetry potentials respectively.

\begin{figure} [tb!]
\centering{\includegraphics[width=8.6cm]{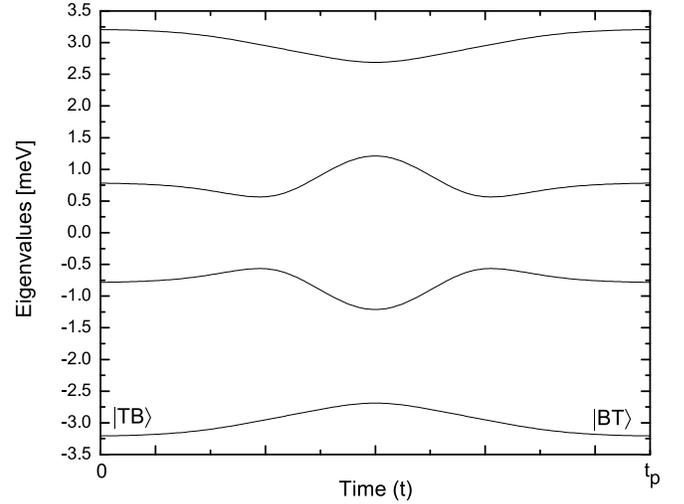}} \caption
{Eigenspectrum for a four donor cell as the pulse scheme given in Eq.~(\ref{eq:EB})~and~(\ref{eq:ES}) is applied to the barrier and symmetry gates respectively, where $S_{\rm{max}}=2\rm{meV}$ and $B_{\rm{max}}=1.2\mathrm{meV}$.\label{fig:ex_gauss}}
\end{figure}

The resulting eigenspectrum is shown for this pulse sequence in Fig.~\ref{fig:ex_gauss}, as a fraction of the pulse time $t_p$.  The degeneracy of the first two states is lifted but the computational states are still strongly populated ($>99.999\%$) before and after the application of the pulse scheme as $E_B\approx0$ at $t=0$ and $t=t_p$.

\section{Time dependant behaviour and the effect of dephasing}
\begin{figure} [tb!]
\centering{\includegraphics[width=8.6cm]{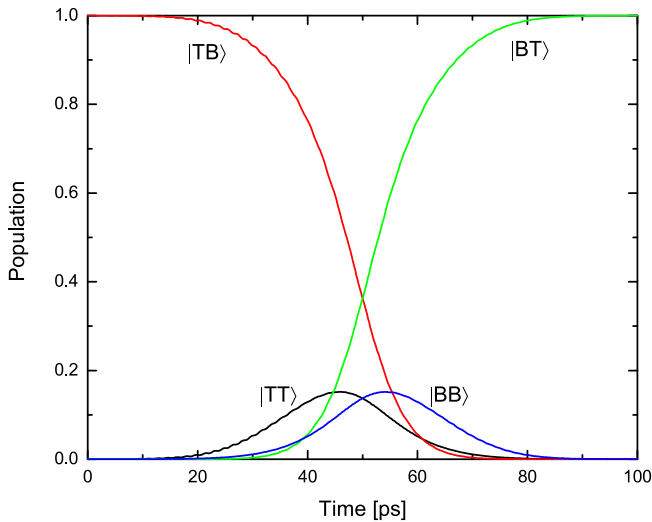}} \caption
{(Color online) Population of states as a function of time showing complete population transfer from the state $|TB\rangle$ to state $|BT\rangle$ while only transiently populating the non-computational states ($|BB\rangle$ and $|TT\rangle$).  The time over which the pulse sequence is applied is $t_p=100 \mathrm{ps}$ and the effects of decoherence are ignored.\label{fig:poptrans}}
\end{figure}

As we are only considering a relatively small basis of states, we solve the density matrix master equation, Eq.~(\ref{eq:mastereq}), to study the time dependence of the system including decoherence.  The equation of motion is
\begin{equation}\label{eq:mastereq}
\dot{\rho}=-\frac{i}{\hbar}\left[H,\rho\right]+\mathcal{L}[\rho],
\end{equation}
where the Liouvillian ($\mathcal{L}[\rho]$) describes the decoherence of the system.  Integrating Eq.~(\ref{eq:mastereq}) in the limit of no decoherence ($\mathcal{L}[\rho]=0$) gives the pure state of the system as it evolves over time.  Fig.~\ref{fig:poptrans} shows the state population for the pulse sequence given in section \ref{sec:CoherentEvol} over a pulse time $t_p=100 \mathrm{ps}$.  The system is initially in $|TB\rangle$ and is then adiabatically switched to $|BT\rangle$ while only transiently occupying the non-computational states.

\begin{figure} [tb!]
\centering{\includegraphics[width=8.6cm]{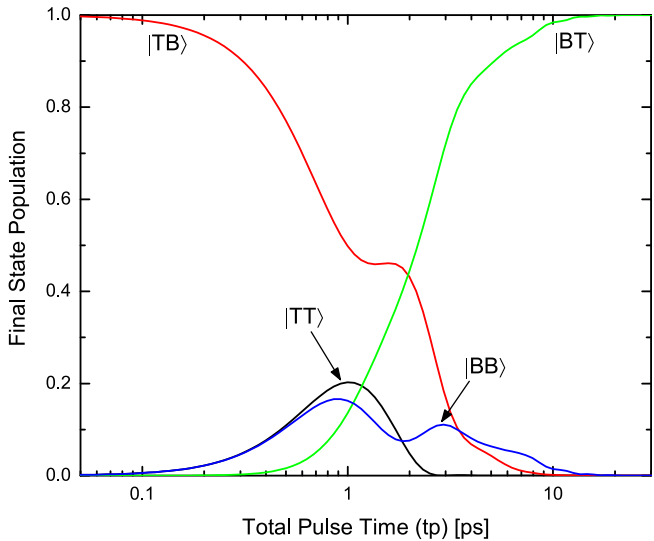}} \caption
{(Color online) Final population of states as a function of total pulse time ($t_p$), ignoring the effects of decoherence. High fidelity transfer ($\ge\!99.95\%$) between computational states is observed for pulse times greater than $20 \rm{ps}$.  For pulse times of less than $0.1 \rm{ps}$ the system does not have time to evolve from its initial state. Between these times, the non-computational states are partially occupied.\label{fig:AdiabaticSweep}}
\end{figure}

To determine the fidelity of transfer, we plot the final occupation probability of each state as a function of total pulse time ($t_p$) assuming we start in state $|TB\rangle$, Fig.~\ref{fig:AdiabaticSweep}.  Three distinct regions can be identified.  For pulse times of less than $0.1\mathrm{ps}$, the pulse sequence is applied too quickly for the system to evolve, which is to be expected as the pulse time is much less than the tunnelling rates of the system.  Pulse times of greater than $20\mathrm{ps}$ satisfy the adiabatic criteria, Eq.~(\ref{eq:AdiabaticEq}), and the system moves smoothly from one computational state to the other with a fidelity of $\ge\!99.95\%$.  Between these regions we see that after switching the non-computational states are occupied with varying probabilities.

\begin{figure} [tb!]
\centering{\includegraphics[width=8.6cm]{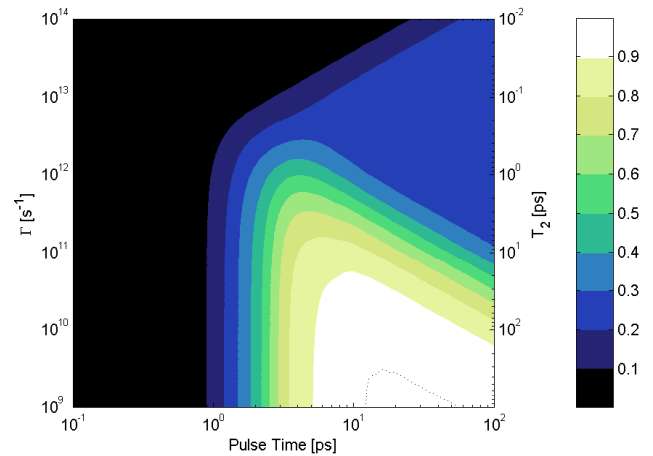}} \caption
{(Color online) Probability of successful transfer as a function of both total pulse time and dephasing time.  The region of $\ge\!99\%$ successful transfer is enclosed by the dotted contour line in the bottom right corner.\label{fig:popwitht2}}
\end{figure}

To study the effects of decoherence, we introduce $\mathcal{L}[\rho] \neq 0$ in Eq.~(\ref{eq:mastereq}).  As we have shown that the relaxation rate ($1/T_1$) due to phonons is expected to be of the order of microseconds, we will only consider a phenomenological $\Gamma_2=1/T_2$ (pure dephasing) rate.  We model this as a decay of the off-diagonal terms of the density matrix, 
\begin{equation}
\mathcal{L}[\rho]=\Gamma_2[\rho-\rm{diag}(\rho)].
\end{equation}
Fig.~\ref{fig:popwitht2} shows the probability of successful transfer from one computational state to the other as a function of total pulse time and dephasing time $T_2$.  The region of $\ge\!99\%$ transfer is enclosed by the dotted contour line on the plot.  The region in the top right corner corresponds to the system dephasing faster than it is being switched, resulting in a loss of fidelity.  The final state in this region is a uniform mixture of the computational and non-computational states.

From Fig.~\ref{fig:popwitht2} we see that even with finite dephasing, there is a window within which coherent transfer can still occur.  For dephasing of $500\mathrm{ps}$ and a total pulse time of $20\mathrm{ps}$, a transfer probability of $>\!99\%$ can be achieved.

\section{Scalability of the buried donor scheme}
The buried donor scheme can be scaled by adding more pairs of donors in a similar fashion to that used for the quantum-dot system to form a line of cells.  The incoherent switching time for a line of cells is predicted to scale approximately linearly based on simulations of quantum-dot systems\cite{Lent:97}.  To compare this to the scaling of the coherent scheme discussed in section~\ref{sec:CoherentEvol}, we estimate the time to adiabatically switch a chain of buried-donor cells.  As shown in Fig.~\ref{fig:QCCA_layout}, this configuration involves a `strip' barrier gate running the length of the chain and a pair of symmetry gates at one end.  The switching of the chain is achieved by applying the same pulse sequence given earlier, Eq.~(\ref{eq:EB})~and~(\ref{eq:ES}), to coherently follow the adiabatic path from one computational state to the other.  This is incorporated into an effective Hamiltonian of the form
\begin{equation}\label{eq:ScalingH}
	H_{\rm{eff}}=E_S(t)\sigma_1^z+E_B(t)\sum_{i=1}^N \sigma^x_i+J_{\rm{ZZ}}\sum_{i=1}^{N-1} \sigma^z_i\sigma^z_{i+1},
\end{equation}
where $N$ is the number of donor pairs.  The interaction term $J_{\rm{ZZ}}$ is approximated using Eq.~(\ref{eq:G2E}) where $J_{\rm{ZZ}}=(E_{\rm{ex}}-E_{\rm{gs}})/2$ for a given cell size $R$.  This model assumes that the entire chain forms a pure state throughout the transfer, which must be performed within the dephasing time of the system.  In this case the minimum evolution time will increase with the number of donor pairs and is controlled by the scaling behaviour of the energy gap between the ground state and the first excited state.  The energy gap is limited by the height of the potential barrier the electron sees which is controlled by $E_B$.  

\begin{figure} [tb!]
\centering{\includegraphics[width=8.6cm]{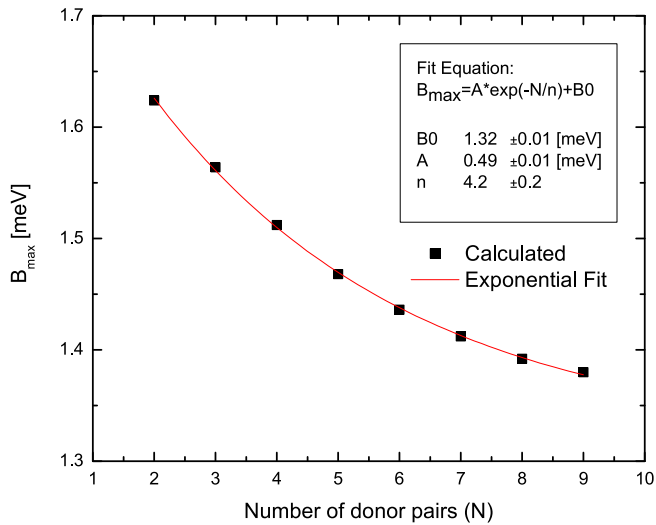}} \caption
{Scaling behaviour for the maximum barrier coupling ($B_{\mathrm{max}}$) which can be applied while still maintaining the minimum energy gap at the centre of the eigenspectrum (the degeneracy point).  $B_{\mathrm{max}}$ is calculated for increasing numbers of donors pairs and then fitted to an exponential function.\label{fig:ExScaling}}
\end{figure}

To simplify the analysis we will ignore decoherence and only consider situations where the minimum energy gap is positioned at the centre of the eigenspectrum (Fig.~\ref{fig:ex_gauss}) where $E_S=0$.  This puts a limit on the maximum barrier coupling ($B_{\rm{max}}$) which can be applied in order to introduce an energy gap.  Fig.~\ref{fig:ExScaling} shows the maximum coupling ($B_{\mathrm{max}}$) which still maintains the minimum energy gap at the centre of the eigenspectrum.
The scaling behaviour of the maximum allowable barrier coupling is fitted to an exponential decay, 
\begin{equation}
B_{\mathrm{max}}\approx0.49\exp(-N/4.2)+1.32,
\end{equation}
where N is the number of donor pairs and $B_{\mathrm{max}}$ is the maximum barrier coupling in $\mathrm{meV}$.  

\begin{figure} [tb!]
\centering{\includegraphics[width=8.6cm]{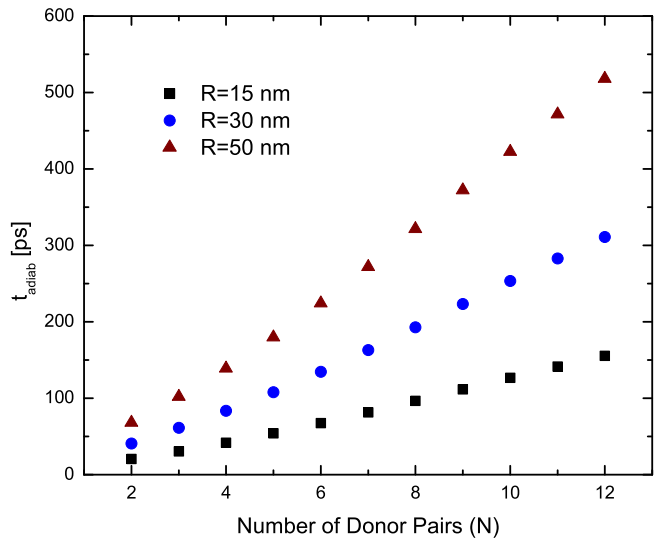}} \caption
{Scaling behaviour of $t_{\mathrm{adiab}}$ as function of number of donor pairs in a BDCA chain for various cell sizes ($R$).  This gives an estimate for the scaling of the minimum allowable evolution time for high fidelity transfer.\label{fig:timescaling}}
\end{figure}

The exact behaviour of the system for large numbers of donors is computationally expensive to calculate.  We can obtain an estimate for the minimum pulse time ($t_p$) which still provides high fidelity ($\ge\!99.95\%$) transfer by observing the scaling of the adiabatic time,
\begin{equation}\label{eq:tadiab}
t_{\mathrm{adiab}}=\kappa\frac{6\sqrt{2}S_{\rm{max}}}{\sqrt{\pi}}\frac{\langle e|\sigma^z_1|g\rangle}{|E_{\mathrm{gap}}|^2},
\end{equation}
with increasing number of donor pairs, where $\kappa$ is a scaling constant used to compare the minimum evolution time with the previous calculations.  This equation is derived by assuming that the adiabatic criteria, Eq.~(\ref{eq:AdiabaticEq}), must be less than $1/\kappa$ to achieve adiabatic evolution and noting that the time derivative of Eq.~(\ref{eq:Controls}) simplifies considerably at the degeneracy point (the centre of the eigenspectrum, Fig.~\ref{fig:ex_gauss}).  At this point the majority of the Hamiltonian is constant in time and the time derivative of the symmetry bias, Eq.~(\ref{eq:ES}), gives the numerical prefactors of Eq.~(\ref{eq:tadiab}).   We use $\kappa=20\rm{ps}$ as this is the required pulse time to achieve $\ge\!99.95\%$ fidelity when switching a single cell comprised of 2 donor pairs with a cell size of $R=15\rm{nm}$.  Fig.~\ref{fig:timescaling} shows $t_{\mathrm{adiab}}$ for up to 12 donor pairs (6 QDCA cells) calculated by diagonalising the effective Hamiltonian of the system, Eq.~(\ref{eq:ScalingH}), for several different cell sizes.  As the cell size is increased, the coupling between pairs of donors is reduced and so the minimum switching time increases.

While $t_{\mathrm{adiab}}$ may underestimate the minimum allowable evolution time, we expect the scaling behaviour of the system to be similar.  The scaling behaviour is approximately linear for these system parameters, though it will ultimately be restricted by the decoherence time of the system.  This could be improved with a stronger bias field or by bringing the donors closer together.

At present, more detailed simulations of logic structures such as those shown in Fig.~\ref{fig:QCAwire} are required to demonstrate the viability of classical processing using BDCA.  Based on simulations of incoherent adiabatic switching\cite{Lent:97}, we expect the coherent control techniques discussed in section~\ref{sec:CoherentEvol} to be applicable to more complex logic structures. 

\section{Application to Solid-State Quantum Computing}

The use of BDCA have a number of applications for solid-state quantum computing, specifically silicon based architectures.  Lines of BDCA cells could be used to initialise charge based qubits and provide an interface between the nano-scale features of the qubits themselves and the micro-scale control electronics.  This would help to shield the qubits from the decoherence effects of the surrounding support electronics.  The use of BDCA lines could also provide an on-chip pathway to transfer classical information between nodes in a solid-state implementation of a semi-classical or type-II quantum computer\cite{Yepez:01}. 

The adiabatic evolution discussed here could be applied to many types of coherent systems based on quantum-dots, not just those based on buried donors.  The adiabatic pathway allows for fast switching and high fidelity with minimal requirements on gate timing and accuracy and in this way is similar to Coherent Transfer by Adiabatic Passage\cite{Greentree:04}.  The construction of an array of coherent BDCA cells would also provide a demonstration of the scalability of semiconductor quantum computing schemes.

The advantages of using adiabatic evolution (rather than coherent oscillations) to measure the decoherence properties of a charge qubit has been investigated by Barrett and Milburn\cite{Barrett:03}.  The use of BDCA cells for this type of experiment has other advantages as well.  A line of BDCA cells arranged with a series of sensitive electrometers along its length would enable the switching and decoherence properties of the system to be more accurately measured.  As there is more than one electron moving during a switching cycle, the use of correlated measurements between all of the electrometers (in the style of those used recently for detecting single electron transfer within a double-dot structure\cite{Buehler:03,Buehler:04}) would provide a more accurate measurement than measuring the movement of just one electron.  

\section{Conclusion}
The use of buried dopants has been investigated as a possible implementation of quantum-dot cellular automata, specifically for the case of phosphorous donors in silicon.  The time and energy scales for this system were investigated and a model developed to describe the system evolution.  

For operating temperatures of less than $1K$, the ground state occupation was found to be approximately $100\%$.  The incoherent switching time for a BDCA cell was found to be slow (of order microseconds) compared to other QDCA architectures due to the lack of a defined electron tunnelling pathway and poor coupling between cells.  The use of coherent evolution to provide fast and controllable switching of BDCA cells was investigated for the case where quantum coherence can be maintained throughout the switching process.  This was found to provide a fast and effective switching mechanism with a cell of side length $15\rm{nm}$ having a coherent switching time of $20 \rm{ps}$ and a fidelity of greater than $99.95\%$.   

The effects of dephasing on this process was investigated and found to have minimal effect as long as the dephasing time is approximately 10-100 times greater than the switching time.  The scaling behaviour of the system was investigated for a simple line of cells and found to scale approximately linearly with the number of cells.  

\begin{acknowledgments}
This work was supported in part by the Australian Research Council, the Australian government, the US National Security Agency, the Advanced Research and Development Activity and the US Army Research Office under contract number DAAD19-01-1-0653.
\end{acknowledgments}

\newpage 
\bibliography{bdqca}

\begin{thebibliography}{31}
\expandafter\ifx\csname natexlab\endcsname\relax\def\natexlab#1{#1}\fi
\expandafter\ifx\csname bibnamefont\endcsname\relax
  \def\bibnamefont#1{#1}\fi
\expandafter\ifx\csname bibfnamefont\endcsname\relax
  \def\bibfnamefont#1{#1}\fi
\expandafter\ifx\csname citenamefont\endcsname\relax
  \def\citenamefont#1{#1}\fi
\expandafter\ifx\csname url\endcsname\relax
  \def\url#1{\texttt{#1}}\fi
\expandafter\ifx\csname urlprefix\endcsname\relax\def\urlprefix{URL }\fi
\providecommand{\bibinfo}[2]{#2}
\providecommand{\eprint}[2][]{\url{#2}}

\bibitem[{\citenamefont{Tougaw and Lent}(1994)}]{Tougaw:94}
\bibinfo{author}{\bibfnamefont{P.~D.} \bibnamefont{Tougaw}} \bibnamefont{and}
  \bibinfo{author}{\bibfnamefont{C.~S.} \bibnamefont{Lent}},
  \bibinfo{journal}{J Appl. Phys.} \textbf{\bibinfo{volume}{75}},
  \bibinfo{pages}{1818} (\bibinfo{year}{1994}).

\bibitem[{\citenamefont{Smith}(1999)}]{Smith:99}
\bibinfo{author}{\bibfnamefont{C.~G.} \bibnamefont{Smith}},
  \bibinfo{journal}{Science} \textbf{\bibinfo{volume}{284}},
  \bibinfo{pages}{274} (\bibinfo{year}{1999}).

\bibitem[{\citenamefont{Timler and Lent}(2003)}]{Timler:03}
\bibinfo{author}{\bibfnamefont{J.}~\bibnamefont{Timler}} \bibnamefont{and}
  \bibinfo{author}{\bibfnamefont{C.~S.} \bibnamefont{Lent}},
  \bibinfo{journal}{J. Appl. Phys.} \textbf{\bibinfo{volume}{94}},
  \bibinfo{pages}{1050} (\bibinfo{year}{2003}).

\bibitem[{\citenamefont{Orlov et~al.}(1997)\citenamefont{Orlov, Amlani,
  Bernstein, Lent, and Snider}}]{Orlov:97}
\bibinfo{author}{\bibfnamefont{A.~O.} \bibnamefont{Orlov}},
  \bibinfo{author}{\bibfnamefont{I.}~\bibnamefont{Amlani}},
  \bibinfo{author}{\bibfnamefont{G.~H.} \bibnamefont{Bernstein}},
  \bibinfo{author}{\bibfnamefont{C.~S.} \bibnamefont{Lent}}, \bibnamefont{and}
  \bibinfo{author}{\bibfnamefont{G.~L.} \bibnamefont{Snider}},
  \bibinfo{journal}{Science} \textbf{\bibinfo{volume}{277}},
  \bibinfo{pages}{928} (\bibinfo{year}{1997}).

\bibitem[{\citenamefont{Orlov et~al.}(1999)\citenamefont{Orlov, Amlani, Toth,
  Lent, Bernstein, and Snider}}]{Orlov:99}
\bibinfo{author}{\bibfnamefont{A.~O.} \bibnamefont{Orlov}},
  \bibinfo{author}{\bibfnamefont{I.}~\bibnamefont{Amlani}},
  \bibinfo{author}{\bibfnamefont{G.}~\bibnamefont{Toth}},
  \bibinfo{author}{\bibfnamefont{C.~S.} \bibnamefont{Lent}},
  \bibinfo{author}{\bibfnamefont{G.~H.} \bibnamefont{Bernstein}},
  \bibnamefont{and} \bibinfo{author}{\bibfnamefont{G.~L.}
  \bibnamefont{Snider}}, \bibinfo{journal}{Appl. Phys. Lett.}
  \textbf{\bibinfo{volume}{74}}, \bibinfo{pages}{2875} (\bibinfo{year}{1999}).

\bibitem[{\citenamefont{Gardelis et~al.}(2003)\citenamefont{Gardelis, Smith,
  Cooper, Ritchie, Linfield, and Jin}}]{Gardelis:03}
\bibinfo{author}{\bibfnamefont{S.}~\bibnamefont{Gardelis}},
  \bibinfo{author}{\bibfnamefont{C.~G.} \bibnamefont{Smith}},
  \bibinfo{author}{\bibfnamefont{J.}~\bibnamefont{Cooper}},
  \bibinfo{author}{\bibfnamefont{D.~A.} \bibnamefont{Ritchie}},
  \bibinfo{author}{\bibfnamefont{E.~H.} \bibnamefont{Linfield}},
  \bibnamefont{and} \bibinfo{author}{\bibfnamefont{Y.}~\bibnamefont{Jin}},
  \bibinfo{journal}{Phys. Rev. B} \textbf{\bibinfo{volume}{67}},
  \bibinfo{pages}{33302} (\bibinfo{year}{2003}).

\bibitem[{\citenamefont{Amlani et~al.}(1999)\citenamefont{Amlani, Orlov, Toth,
  Bernstein, Lent, and Snider}}]{Amlani:99}
\bibinfo{author}{\bibfnamefont{I.}~\bibnamefont{Amlani}},
  \bibinfo{author}{\bibfnamefont{A.~O.} \bibnamefont{Orlov}},
  \bibinfo{author}{\bibfnamefont{G.}~\bibnamefont{Toth}},
  \bibinfo{author}{\bibfnamefont{G.~H.} \bibnamefont{Bernstein}},
  \bibinfo{author}{\bibfnamefont{C.~S.} \bibnamefont{Lent}}, \bibnamefont{and}
  \bibinfo{author}{\bibfnamefont{G.~L.} \bibnamefont{Snider}},
  \bibinfo{journal}{Science} \textbf{\bibinfo{volume}{284}},
  \bibinfo{pages}{289} (\bibinfo{year}{1999}).

\bibitem[{\citenamefont{Toth and Lent}(2001)}]{Toth:01}
\bibinfo{author}{\bibfnamefont{G.}~\bibnamefont{Toth}} \bibnamefont{and}
  \bibinfo{author}{\bibfnamefont{C.~S.} \bibnamefont{Lent}},
  \bibinfo{journal}{Phys. Rev. A.} \textbf{\bibinfo{volume}{63}},
  \bibinfo{pages}{052315} (\bibinfo{year}{2001}).

\bibitem[{\citenamefont{Oi et~al.}(Unpublished)\citenamefont{Oi, Greentree, and
  Schirmer}}]{Oi:04}
\bibinfo{author}{\bibfnamefont{D.~K.~L.} \bibnamefont{Oi}},
  \bibinfo{author}{\bibfnamefont{A.~D.} \bibnamefont{Greentree}},
  \bibnamefont{and} \bibinfo{author}{\bibfnamefont{S.~G.}
  \bibnamefont{Schirmer}} (\bibinfo{year}{Unpublished}),
  \eprint{arXiv:quant-ph/0412122}.

\bibitem[{\citenamefont{Hollenberg et~al.}(2004)\citenamefont{Hollenberg,
  Dzurak, Wellard, Hamilton, Reilly, Milburn, and Clark}}]{Hollenberg:04}
\bibinfo{author}{\bibfnamefont{L.~C.~L.} \bibnamefont{Hollenberg}},
  \bibinfo{author}{\bibfnamefont{A.~S.} \bibnamefont{Dzurak}},
  \bibinfo{author}{\bibfnamefont{C.}~\bibnamefont{Wellard}},
  \bibinfo{author}{\bibfnamefont{A.~R.} \bibnamefont{Hamilton}},
  \bibinfo{author}{\bibfnamefont{D.~J.} \bibnamefont{Reilly}},
  \bibinfo{author}{\bibfnamefont{G.~J.} \bibnamefont{Milburn}},
  \bibnamefont{and} \bibinfo{author}{\bibfnamefont{R.~G.} \bibnamefont{Clark}},
  \bibinfo{journal}{Phys. Rev. B} \textbf{\bibinfo{volume}{69}},
  \bibinfo{pages}{113301} (\bibinfo{year}{2004}).

\bibitem[{\citenamefont{Dzurak et~al.}(Unpublished)\citenamefont{Dzurak,
  Hollenberg, Jamieson, Stanley, Yang, Buehler, Chan, Reilly, Wellard, Hamilton
  et~al.}}]{Dzurak:03}
\bibinfo{author}{\bibfnamefont{A.~S.} \bibnamefont{Dzurak}},
  \bibinfo{author}{\bibfnamefont{L.}~\bibnamefont{Hollenberg}},
  \bibinfo{author}{\bibfnamefont{D.~N.} \bibnamefont{Jamieson}},
  \bibinfo{author}{\bibfnamefont{F.~E.} \bibnamefont{Stanley}},
  \bibinfo{author}{\bibfnamefont{C.}~\bibnamefont{Yang}},
  \bibinfo{author}{\bibfnamefont{T.~M.} \bibnamefont{Buehler}},
  \bibinfo{author}{\bibfnamefont{V.}~\bibnamefont{Chan}},
  \bibinfo{author}{\bibfnamefont{D.~J.} \bibnamefont{Reilly}},
  \bibinfo{author}{\bibfnamefont{C.}~\bibnamefont{Wellard}},
  \bibinfo{author}{\bibfnamefont{A.~R.} \bibnamefont{Hamilton}},
  \bibnamefont{et~al.} (\bibinfo{year}{Unpublished}),
  \eprint{arXiv:cond-mat/0306265}.

\bibitem[{\citenamefont{Schofield et~al.}(2003)\citenamefont{Schofield, Curson,
  Simmons, Ruess, Hallam, Oberbeck, and Clark}}]{Schofield:03}
\bibinfo{author}{\bibfnamefont{S.~R.} \bibnamefont{Schofield}},
  \bibinfo{author}{\bibfnamefont{N.~J.} \bibnamefont{Curson}},
  \bibinfo{author}{\bibfnamefont{M.~Y.} \bibnamefont{Simmons}},
  \bibinfo{author}{\bibfnamefont{F.~J.} \bibnamefont{Ruess}},
  \bibinfo{author}{\bibfnamefont{T.}~\bibnamefont{Hallam}},
  \bibinfo{author}{\bibfnamefont{L.}~\bibnamefont{Oberbeck}}, \bibnamefont{and}
  \bibinfo{author}{\bibfnamefont{R.~G.} \bibnamefont{Clark}},
  \bibinfo{journal}{Phys. Rev. Lett.} \textbf{\bibinfo{volume}{91}},
  \bibinfo{pages}{136104} (\bibinfo{year}{2003}).

\bibitem[{\citenamefont{Tucker}(2001)}]{Tucker:01}
\bibinfo{author}{\bibfnamefont{J.~R.} \bibnamefont{Tucker}},
  \bibinfo{journal}{Quantum Inf. Comput.} \textbf{\bibinfo{volume}{1}},
  \bibinfo{pages}{129} (\bibinfo{year}{2001}).

\bibitem[{\citenamefont{Schenkel et~al.}(2003)\citenamefont{Schenkel, Persaud,
  Park, Nilsson, Bokor, Liddle, Keller, Schneider, Cheng, and
  Humphries}}]{Schenkel:03}
\bibinfo{author}{\bibfnamefont{T.}~\bibnamefont{Schenkel}},
  \bibinfo{author}{\bibfnamefont{A.}~\bibnamefont{Persaud}},
  \bibinfo{author}{\bibfnamefont{S.~J.} \bibnamefont{Park}},
  \bibinfo{author}{\bibfnamefont{J.}~\bibnamefont{Nilsson}},
  \bibinfo{author}{\bibfnamefont{J.}~\bibnamefont{Bokor}},
  \bibinfo{author}{\bibfnamefont{J.~A.} \bibnamefont{Liddle}},
  \bibinfo{author}{\bibfnamefont{R.}~\bibnamefont{Keller}},
  \bibinfo{author}{\bibfnamefont{D.~H.} \bibnamefont{Schneider}},
  \bibinfo{author}{\bibfnamefont{D.~W.} \bibnamefont{Cheng}}, \bibnamefont{and}
  \bibinfo{author}{\bibfnamefont{D.~E.} \bibnamefont{Humphries}},
  \bibinfo{journal}{J Appl. Phys.} \textbf{\bibinfo{volume}{94}},
  \bibinfo{pages}{7017} (\bibinfo{year}{2003}).

\bibitem[{\citenamefont{Clark et~al.}(2003)\citenamefont{Clark, Brenner,
  Buehler, Chan, Curson, Dzurak, Gauja, Goan, Greentree, Hallam
  et~al.}}]{Clark:03}
\bibinfo{author}{\bibfnamefont{R.~G.} \bibnamefont{Clark}},
  \bibinfo{author}{\bibfnamefont{R.}~\bibnamefont{Brenner}},
  \bibinfo{author}{\bibfnamefont{T.~M.} \bibnamefont{Buehler}},
  \bibinfo{author}{\bibfnamefont{V.}~\bibnamefont{Chan}},
  \bibinfo{author}{\bibfnamefont{N.~J.} \bibnamefont{Curson}},
  \bibinfo{author}{\bibfnamefont{A.~S.} \bibnamefont{Dzurak}},
  \bibinfo{author}{\bibfnamefont{E.}~\bibnamefont{Gauja}},
  \bibinfo{author}{\bibfnamefont{H.~S.} \bibnamefont{Goan}},
  \bibinfo{author}{\bibfnamefont{A.~D.} \bibnamefont{Greentree}},
  \bibinfo{author}{\bibfnamefont{T.}~\bibnamefont{Hallam}},
  \bibnamefont{et~al.}, \bibinfo{journal}{Philos. Trans. R. Soc. Lond. Ser.
  A-Math. Phys. Eng. Sci.} \textbf{\bibinfo{volume}{361}},
  \bibinfo{pages}{1451} (\bibinfo{year}{2003}).

\bibitem[{\citenamefont{Snider et~al.}(1999)\citenamefont{Snider, Orlov,
  Amlani, Zuo, Bernstein, Lent, Merz, and Porod}}]{Snider:99b}
\bibinfo{author}{\bibfnamefont{G.~L.} \bibnamefont{Snider}},
  \bibinfo{author}{\bibfnamefont{A.~O.} \bibnamefont{Orlov}},
  \bibinfo{author}{\bibfnamefont{I.}~\bibnamefont{Amlani}},
  \bibinfo{author}{\bibfnamefont{X.}~\bibnamefont{Zuo}},
  \bibinfo{author}{\bibfnamefont{G.~H.} \bibnamefont{Bernstein}},
  \bibinfo{author}{\bibfnamefont{C.~S.} \bibnamefont{Lent}},
  \bibinfo{author}{\bibfnamefont{J.~L.} \bibnamefont{Merz}}, \bibnamefont{and}
  \bibinfo{author}{\bibfnamefont{W.}~\bibnamefont{Porod}}, \bibinfo{journal}{J
  Vac. Sci. Technol. A} \textbf{\bibinfo{volume}{17}}, \bibinfo{pages}{1394}
  (\bibinfo{year}{1999}).

\bibitem[{\citenamefont{Kohn}(1955)}]{Kohn:55}
\bibinfo{author}{\bibfnamefont{W.}~\bibnamefont{Kohn}}, in
  \emph{\bibinfo{booktitle}{Solid state physics}}, edited by
  \bibinfo{editor}{\bibfnamefont{F.}~\bibnamefont{Seitz}} \bibnamefont{and}
  \bibinfo{editor}{\bibfnamefont{D.}~\bibnamefont{Turnbull}}
  (\bibinfo{publisher}{Academic Press}, \bibinfo{address}{New York},
  \bibinfo{year}{1955}), vol.~\bibinfo{volume}{5}, pp.
  \bibinfo{pages}{257--320}.

\bibitem[{\citenamefont{Slater}(1963)}]{Slater:63}
\bibinfo{author}{\bibfnamefont{J.~C.} \bibnamefont{Slater}},
  \emph{\bibinfo{title}{Quantum theory of molecules and solids}},
  vol.~\bibinfo{volume}{1} of \emph{\bibinfo{series}{International series in
  pure and applied physics.}} (\bibinfo{publisher}{McGraw-Hill},
  \bibinfo{address}{New York}, \bibinfo{year}{1963}).

\bibitem[{\citenamefont{Appel}(1964)}]{Appel:64}
\bibinfo{author}{\bibfnamefont{J.}~\bibnamefont{Appel}},
  \bibinfo{journal}{Phys. Rev.} \textbf{\bibinfo{volume}{133}},
  \bibinfo{pages}{A280} (\bibinfo{year}{1964}).

\bibitem[{\citenamefont{Grabert and Devoret}(1992)}]{Grabert:92}
\bibinfo{author}{\bibfnamefont{H.}~\bibnamefont{Grabert}} \bibnamefont{and}
  \bibinfo{author}{\bibfnamefont{M.~H.} \bibnamefont{Devoret}},
  \emph{\bibinfo{title}{Single charge tunneling : Coulomb blockade phenomena in
  nanostructures}} (\bibinfo{publisher}{Plenum Press}, \bibinfo{address}{New
  York}, \bibinfo{year}{1992}).

\bibitem[{\citenamefont{Lee et~al.}(2005)\citenamefont{Lee, Greentree, Dinale,
  Escott, Dzurak, and Clark}}]{Lee:04}
\bibinfo{author}{\bibfnamefont{K.~H.} \bibnamefont{Lee}},
  \bibinfo{author}{\bibfnamefont{A.~D.} \bibnamefont{Greentree}},
  \bibinfo{author}{\bibfnamefont{J.~P.} \bibnamefont{Dinale}},
  \bibinfo{author}{\bibfnamefont{C.~C.} \bibnamefont{Escott}},
  \bibinfo{author}{\bibfnamefont{A.~S.} \bibnamefont{Dzurak}},
  \bibnamefont{and} \bibinfo{author}{\bibfnamefont{R.~G.} \bibnamefont{Clark}},
  \bibinfo{journal}{Nanotechnology} \textbf{\bibinfo{volume}{16}},
  \bibinfo{pages}{7} (\bibinfo{year}{2005}).

\bibitem[{\citenamefont{Barrett and Milburn}(2003)}]{Barrett:03}
\bibinfo{author}{\bibfnamefont{S.~D.} \bibnamefont{Barrett}} \bibnamefont{and}
  \bibinfo{author}{\bibfnamefont{G.~J.} \bibnamefont{Milburn}},
  \bibinfo{journal}{Phys. Rev. B} \textbf{\bibinfo{volume}{68}},
  \bibinfo{pages}{155307} (\bibinfo{year}{2003}).

\bibitem[{\citenamefont{Fedichkin and Fedorov}(2004)}]{Fedichkin:04}
\bibinfo{author}{\bibfnamefont{L.}~\bibnamefont{Fedichkin}} \bibnamefont{and}
  \bibinfo{author}{\bibfnamefont{A.}~\bibnamefont{Fedorov}},
  \bibinfo{journal}{Phys. Rev. A} \textbf{\bibinfo{volume}{69}},
  \bibinfo{pages}{032311} (\bibinfo{year}{2004}).

\bibitem[{\citenamefont{Bockelmann and Bastard}(1990)}]{Bockelmann:90}
\bibinfo{author}{\bibfnamefont{U.}~\bibnamefont{Bockelmann}} \bibnamefont{and}
  \bibinfo{author}{\bibfnamefont{G.}~\bibnamefont{Bastard}},
  \bibinfo{journal}{Phys. Rev. B} \textbf{\bibinfo{volume}{42}},
  \bibinfo{pages}{8947} (\bibinfo{year}{1990}).

\bibitem[{\citenamefont{Vitanov et~al.}(2001)\citenamefont{Vitanov, Halfmann,
  Shore, and Bergmann}}]{Vitanov:01}
\bibinfo{author}{\bibfnamefont{N.~V.} \bibnamefont{Vitanov}},
  \bibinfo{author}{\bibfnamefont{T.}~\bibnamefont{Halfmann}},
  \bibinfo{author}{\bibfnamefont{B.~W.} \bibnamefont{Shore}}, \bibnamefont{and}
  \bibinfo{author}{\bibfnamefont{K.}~\bibnamefont{Bergmann}},
  \bibinfo{journal}{Annu. Rev. Phys. Chem.} \textbf{\bibinfo{volume}{52}},
  \bibinfo{pages}{763} (\bibinfo{year}{2001}).

\bibitem[{\citenamefont{Lent and Tougaw}(1997)}]{Lent:97}
\bibinfo{author}{\bibfnamefont{C.}~\bibnamefont{Lent}} \bibnamefont{and}
  \bibinfo{author}{\bibfnamefont{P.}~\bibnamefont{Tougaw}},
  \bibinfo{journal}{Proc. IEEE} \textbf{\bibinfo{volume}{85}},
  \bibinfo{pages}{541} (\bibinfo{year}{1997}).

\bibitem[{\citenamefont{Messiah}(1965)}]{Messiah:65}
\bibinfo{author}{\bibfnamefont{A.}~\bibnamefont{Messiah}},
  \emph{\bibinfo{title}{Quantum mechanics}}, vol.~\bibinfo{volume}{2}
  (\bibinfo{publisher}{North-Holland}, \bibinfo{address}{Amsterdam},
  \bibinfo{year}{1965}).

\bibitem[{\citenamefont{Yepez}(2001)}]{Yepez:01}
\bibinfo{author}{\bibfnamefont{J.}~\bibnamefont{Yepez}}, \bibinfo{journal}{Int.
  J. Mod. Phys. C} \textbf{\bibinfo{volume}{12}}, \bibinfo{pages}{1273}
  (\bibinfo{year}{2001}).

\bibitem[{\citenamefont{Greentree et~al.}(2004)\citenamefont{Greentree, Cole,
  Hamilton, and Hollenberg}}]{Greentree:04}
\bibinfo{author}{\bibfnamefont{A.~D.} \bibnamefont{Greentree}},
  \bibinfo{author}{\bibfnamefont{J.~H.} \bibnamefont{Cole}},
  \bibinfo{author}{\bibfnamefont{A.~R.} \bibnamefont{Hamilton}},
  \bibnamefont{and} \bibinfo{author}{\bibfnamefont{L.~C.~L.}
  \bibnamefont{Hollenberg}}, \bibinfo{journal}{Phys. Rev. B}
  \textbf{\bibinfo{volume}{70}}, \bibinfo{pages}{235317}
  (\bibinfo{year}{2004}).

\bibitem[{\citenamefont{Buehler et~al.}(2003)\citenamefont{Buehler, Reilly,
  Brenner, Hamilton, Dzurak, and Clark}}]{Buehler:03}
\bibinfo{author}{\bibfnamefont{T.~M.} \bibnamefont{Buehler}},
  \bibinfo{author}{\bibfnamefont{D.~J.} \bibnamefont{Reilly}},
  \bibinfo{author}{\bibfnamefont{R.}~\bibnamefont{Brenner}},
  \bibinfo{author}{\bibfnamefont{A.~R.} \bibnamefont{Hamilton}},
  \bibinfo{author}{\bibfnamefont{A.~S.} \bibnamefont{Dzurak}},
  \bibnamefont{and} \bibinfo{author}{\bibfnamefont{R.~G.} \bibnamefont{Clark}},
  \bibinfo{journal}{Appl. Phys. Lett.} \textbf{\bibinfo{volume}{82}},
  \bibinfo{pages}{577} (\bibinfo{year}{2003}).

\bibitem[{\citenamefont{Buehler et~al.}(2004)\citenamefont{Buehler, Reilly,
  Starrett, Court, Hamilton, Dzurak, and Clark}}]{Buehler:04}
\bibinfo{author}{\bibfnamefont{T.~M.} \bibnamefont{Buehler}},
  \bibinfo{author}{\bibfnamefont{D.~J.} \bibnamefont{Reilly}},
  \bibinfo{author}{\bibfnamefont{R.~P.} \bibnamefont{Starrett}},
  \bibinfo{author}{\bibfnamefont{N.~A.} \bibnamefont{Court}},
  \bibinfo{author}{\bibfnamefont{A.~R.} \bibnamefont{Hamilton}},
  \bibinfo{author}{\bibfnamefont{A.~S.} \bibnamefont{Dzurak}},
  \bibnamefont{and} \bibinfo{author}{\bibfnamefont{R.~G.} \bibnamefont{Clark}},
  \bibinfo{journal}{J Appl. Phys.} \textbf{\bibinfo{volume}{96}},
  \bibinfo{pages}{4508} (\bibinfo{year}{2004}).

\end{thebibliography}

\end{document}